\documentclass[reprint,showpacs,floatfix,amsmath,amssymb,prb]{revtex4-2}
\usepackage{graphicx,tikz,latexsym,hyperref,colortbl}
\usepackage{multirow}
\usepgflibrary{shapes.arrows}
\usepgfmodule{decorations} \usepackage{fancyhdr} \usepackage{textcomp}

\newcommand{\nmf}[1]{$\mathrm{Na#1F_3}$}

\newcommand{\fref}[1]{Figure~\ref{fig:#1}}
\newcommand{\tref}[1]{Table~\ref{tab:#1}}

\hypersetup{colorlinks=false, citecolor=purple, linkcolor=red}

\begin{document}
\title{Structural  and   magnetic  characterization  of   the  elusive
  Jahn-Teller active $\mathrm{NaCrF_3}$}

\author{Fabian L. M. Bernal}
\affiliation{Chemistry Department and Center for Material Science and Nanotechnology, University of Oslo, NO-0315, Norway}
\author{\O ystein S. Fjellv\aa g}
\affiliation{Department for Neutron Materials Characterization, Institute for Energy Technology, PO Box 40, NO-2027, Kjeller, Norway}
\author{Jonas Sottmann}
\affiliation{Chemistry Department and Center for Material Science and Nanotechnology, University of Oslo, NO-0315, Norway}
\author{David S. Wragg}
\affiliation{Chemistry Department and Center for Material Science and Nanotechnology, University of Oslo, NO-0315, Norway}
\author{Helmer Fjellv\aa g}
\affiliation{Chemistry Department and Center for Material Science and Nanotechnology, University of Oslo, NO-0315, Norway}
\author{Christina Drathen}
\affiliation{ESRF- The European Synchrotron, 71, Avenue des Martyrs, Grenoble 38043, France}
\affiliation{Current Address: Bruker AXS, {\"O}estliche Rheinbrueckenstr. 49, 76187 Karlsruhe, Germany.}
\author{Wojciech A. S\l awi\'{n}ski}
\affiliation{Faculty of Chemistry, University of Warsaw, Pasteura 1, 02-093 Warsaw, Poland}
\affiliation{ISIS Facility, Rutherford Appleton Laboratory, Harwell Oxford, Didcot, Oxfordshire OX11 0QX, U.K.}
\author{Ole Martin L\o vvik}
\affiliation{Department of Physics, University of Oslo, NO-0315, Norway}

\begin{abstract}
  We report on  the structural and magnetic properties  of the elusive
  Jahn-Teller  active  compound  $\mathrm{NaCrF_3}$,  for  first  time
  synthesized in large  quantities allowing detailed characterization.
  The crystal  structure of $\mathrm{NaCrF_3}$ is  initially described
  from  a DFT  model  which helped  serve as  basis  for indexing  and
  structure  determination  confirmed by  high-resolution  synchrotron
  X-ray  diffraction   experiments.   $\mathrm{NaCrF_3}$   adopts  the
  triclinic  space  group  $P\bar{1}$ (isostructural  with  \nmf{Cu}).
  Magnetometry studies at low temperature show that $\mathrm{NaCrF_3}$
  is  a weak  antiferromagnet, Curie  temperature $\theta=-4$~K.   The
  N\'eel  temperature  is  $T_N=21.3$~K and  the  paramagnetic  moment
  $\mu=4.47\;  \mu_{_B}$ in  accordance  with  the theoretical  $S=2$.
  Field-dependent measurements between 2 and  12 K unveil the onset of
  metamagnetic  behavior.  Our  experiments revealed  a weakly  canted
  $A$-type magnetic structure observed  by neutron powder diffraction,
  with  a magnetic  propagation  vector $(1/2,1/2,0)$  and a  magnetic
  moment of 3.51~$\mu_B$ at 1.5~K.   Our results shed further light on
  the Jahn-Teller  effects and  strong correlations  as a  function of
  $A$-ion size in the family $A\mathrm{CrF_3}$.
\end{abstract}
\maketitle

\begin{figure*}[t!]
  \centering
 \includegraphics[scale=.85]{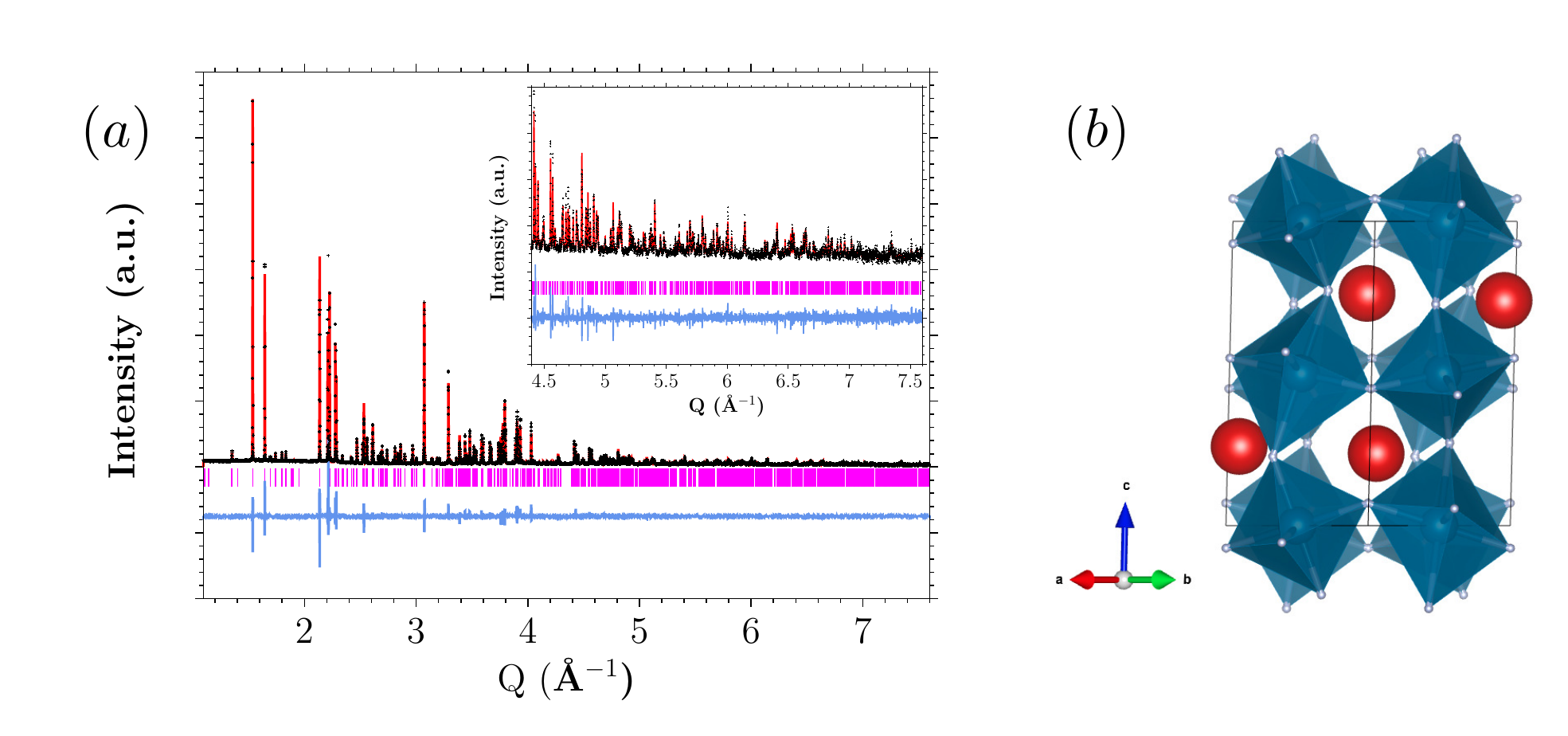}
 \caption{($a$) Final observed (black dots), calculated (red line) and
   difference  (blue   line)  synchrotron  X-ray   powder  diffraction
   profiles ($\lambda =  0.4501$ \AA) for $\mathrm{NaCrF_3}$  at 298 K
   ($a =  5.51515(2)$ \AA, $  b = 5.68817(3)$  \AA, $ c  = 8.18349(3)$
   \AA,  $\alpha =  90.5039(3)^\circ$,  $\beta  = 92.2554(3)  ^\circ$,
   $\gamma    =     86.0599(2)^\circ)$.     $R_{wp}     =    11.52\%$;
   $R_{exp} =  5.53\%$.  Inset, close up  of the high angle  region of
   the refined pattern.  (b) Structure of ${\rm NaCrF_3}$ viewed along
   the [110]-direction.}
  \label{fig:f1}
\end{figure*}

\begin{figure*}[t!]
  \centering
 \includegraphics[scale=.85]{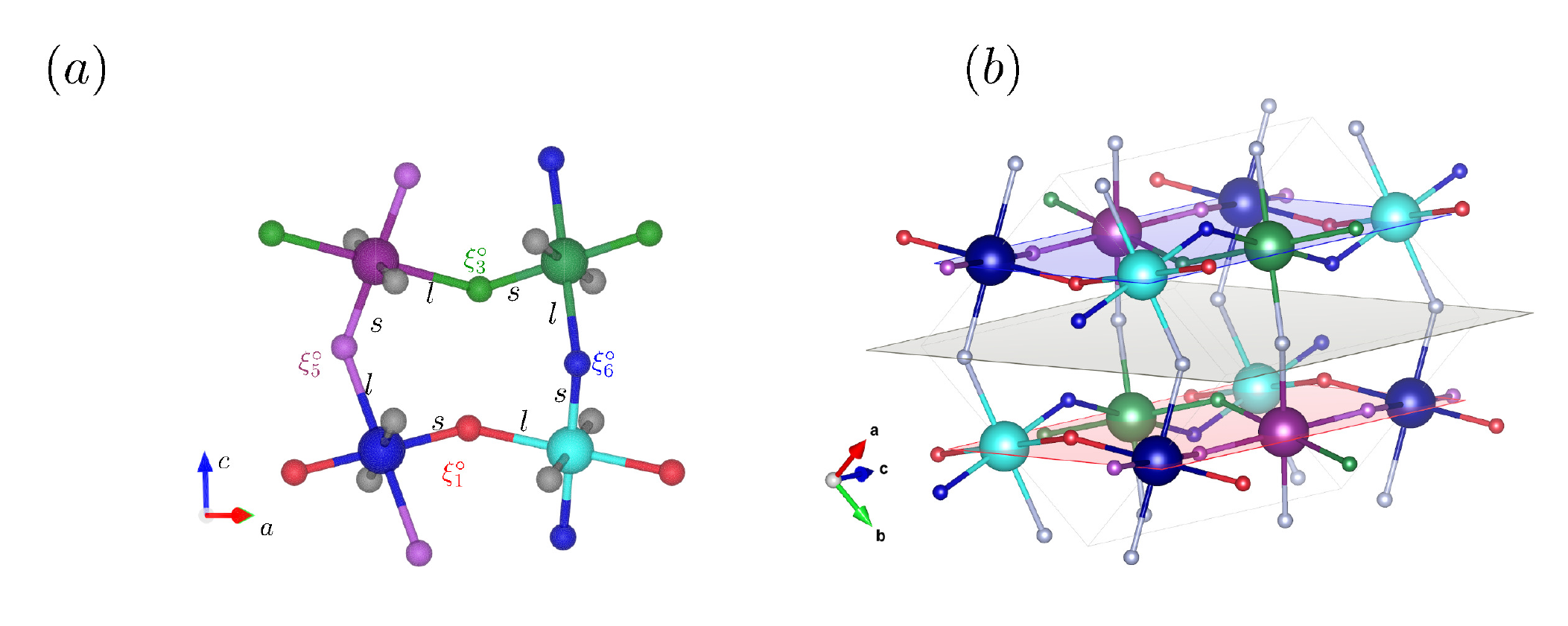}
 \caption{ ($a$) $ls$- bond length  motif of the four crystallographic
   chromium sites  of $\mathrm{NaCrF_3}$.  Cr1,  Cr2, Cr3 and  Cr4 are
   represented  here  with  blue,  cyan,  purple  and  green  spheres,
   respectively.  The  Cr-F-Cr bond angles are  labelled: $\xi^\circ$:
   $\xi^\circ_1=\mathrm{Cr1-F1-Cr2}$,
   $\xi^\circ_6=\mathrm{Cr2-F6-Cr4}$,
   $\xi^\circ_3=\mathrm{Cr4-F3-Cr3}$,
   $\xi^\circ_5=\mathrm{Cr3-F5-Cr1}$.  ($b$)  Packed crystal structure
   of $\mathrm{NaCrF_3}$  with red and  blue planes marking  layers of
   $\mathrm{Cu^{2+}}$ in which the ls-bond length motif (a) is rotated
   by  90$^\circ$ relative  to  the adjacent  layers.   The mid  plane
   (gray)  cuts   through  the  connecting  $m$-bond   distances,  and
   represents the  stacking directions  of the  canted AOO.   The unit
   cell is shown in pale grey.}
  \label{fig:f1b}
\end{figure*}

\section{Introduction}
\label{sec:intro}

The cooperative Jahn-Teller \cite{JT} (JT) effect is commonly ascribed
to   structural   distortions   caused   by   the   coupling   between
electronically degenerate orbital states  of transition metal ions and
their normal modes  of vibration.  This coupling  results in reduction
of the symmetry of the bonding  environment around the JT ion to lower
the total  energy.  JT-  active perovskite-type  materials are  at the
center of intensive research within the material science community for
their  wide range  of  physical properties  and structural  diversity.
Superconductivity,  colossal   magnetoresistance  (CMR)   and  polaron
confinement  are known  for  these compounds,  giving applications  in
information storage  and spintronics  \cite{AN1,AN2,AN3}.  Perovskites
have  the  chemical   formula  $ABX_3  $.   JT-active   ions  such  as
$\mathrm{Mn^{3+},  Cr^{2+}}$  and  $\mathrm{Cu^{2+}}$  (with  electron
configurations 3$d^4$, 3$d^4$ and 3$d^9$, respectively) can occupy the
octahedral  $B$-site (e.g.,  $[\mathrm{MnO}_6]$).   The octahedra  are
linked  by their  vertices  forming sets  of  $B$-$X$-$B$ bond  angles
$\xi^\circ$   (defined   here   as   the   perovskite   angle).    The
electron-phonon  coupling   (ie.,  $E\otimes  e$)   causes  octahedral
distortion  which  favor  the  occupation of  one  of  the  originally
degenerate orbital  states. At  the same time,  the choice  of orbital
state induces an orbital ordering (OO).

The  best  known JT-active  oxide  perovskite  is lanthanum  manganite
$\mathrm{LaMnO_3}$,   a  parent   compound   for  several   derivative
crystalline  compounds exhibiting  CMR.  An  essential feature  of the
manganites is  the role played by  the atom occupying the  $A-$site in
influencing deformations of the perovskite type structure, and thereby
also the  JT-structural distortions,  leading to  a rich  diversity of
spin, orbital  and charge  orderings.  In  fluorides JT-ions  are well
known  for  showing  interesting  phenomena  under  external  stimuli.
Alkali    ternary    manganese    (III)   fluorides    with    formula
$A_x\mathrm{MnF}_{3+x}$   (with  $A=$Na,   K,  Cs)   show  significant
structural diversity, adopting 0-,1- and 2- dimensional vertex-sharing
arrangements of  the octahedral units  depending on the value  of $x$.
\cite{JCP118,PRB76}.

3-dimensional vertex sharing high spin 3$d^4$ electronic configuration
can  form  perovskite-type  fluoride  structures  (fluoroperovskites).
These  include ternary  chromium (II)  fluoroperovskites with  formula
$A\mathrm{CrF_3}$ (where  $A=$ alkali metals).   $\mathrm{KCrF_3}$ has
two   structural-phase    transitions   at    elevated   temperatures:
$I{112/m}\to  I{4/mcm}$ at  250 K  and $I{4/mcm}\to  P{m3m}$ at  973 K
\cite{JACS128,RSC17},  and  theoretical  studies have  associated  the
metal   to    insulator   transition    with   the   onset    of   the
tetragonal-to-cubic  phase  transition   \cite{PRB84}.   In  addition,
$\mathrm{KCrF_3}$  displays  a  rich  magnetic phase  diagram  at  low
temperatures: an incommensurate antiferromagnetic  ordering at 79.5 K,
an  incommensurate-to-commensurate  antifferromagnetic  transition  at
45.8 K, and below 9.5 K  a canted antiferromagnetic ordering with weak
ferromagnetic interactions  \cite{PRB82}. Further studies of  the role
played  by the  $A-$site  in $A\mathrm{CrF_3}$  are currently  lacking
despite the interesting phase  diagram of $\mathrm{KCrF_3}$.  The main
reason for  this is  the lack  of proper  synthetic protocols  for the
reactions   of   $\mathrm{Cr^{2+}}$compounds  with   fluorides.    The
synthesis  of  $\mathrm{NaCrF_3}$  has   until  now  proved  extremely
problematic due to the sensitivity of $\mathrm{Cr^{2+}}$ to oxidation.
None of the synthesis routes described  by Deyrup and Earnshaw {\em et
  al.}  resulted in $\mathrm{NaCrF_3}$\cite{DEYRUP,EARN}.  To the best
of  our   knowledge,  the   only  evidence   of  the   preparation  of
$\mathrm{NaCrF_3}$  was  given  by  the work  of  Vollmer  and  UV-vis
spectroscopy studies performed by Kruger \cite{VOLL,OELKRUG}.  Our new
reliable synthetic  protocol for  $\mathrm{NaCrF_3}$ opens  up further
possibilities  for synthesizing  analogous materials  of interest  for
information storage technologies, with rich states of matter and novel
physical phenomena to appear  in stoichiometric and non-stoichiometric
modifications  of the  $A$-  and $B$  sites  in the  $A\mathrm{CrF_3}$
family.   We  report for  the  first  time  the crystal  and  magnetic
structure  of  the  elusive JT-  active  compound,  $\mathrm{NaCrF_3}$
prepared  by   a  novel  wet-chemistry  method.    These  results  are
complemented by magnetometry studies.

\begin{table*}[th!]
  \centering
  \caption{Structural parameters from  Rietveld refinement of HR-SPXRD
    dataset of $\mathrm{NaCrF_3}$ at ambient conditions. $l,m$ and $s$
    are long, medium and short bond distances, respectively.}
   \label{tab:t1}
  \begin{tabular}{l c c c c c}
    &&&&&\\
    \hline\hline
    &&&&&\\
    \textbf{Space Group}: & $P\bar{1}$&&&&\\
    $a$ &5.51515(2) \AA&&&&$\Delta d_{Cr1}=78.37\times 10^{-4}$ \AA \\
    $b$ &5.68817(3) \AA&&\textbf{Octahedral distortions}:&&$\Delta d_{Cr2}=59.01\times 10^{-4}$ \AA \\
    $c$ &8.18349(3) \AA&&&&$\Delta d_{Cr3}=72.35\times 10^{-4}$ \AA \\
    $\alpha$ & 90.5039(3)$^\circ$&&&&$\Delta d_{Cr4}=76.86\times 10^{-4}$ \AA\\
    $\beta$ & 92.2554(3)$^\circ$&&&\\
    $\gamma$& 86.0599(2)$^\circ$&&&\\
    $V$& 255.915(2) \AA$^3$&&&\\
                &&&&\\
    $R_{wp}, R_{wp-bkg}$&11.5162, 21.573 &&$R_{p}, R_{p-bkg}$&8.6967, 23.7006 &\\
    $R_{exp}, R_{exp-bkg}$&5.5252, 10.3500  &&$\chi^2$ &2.08&\\
    N$^\circ$ of independent parameters &53&&&\\
    Restrains, constrains&0, 3&&&\\
    Rigid bodies        &0&&&\\
        $Z$&4&&&\\
    &\multicolumn{4}{c}{\textbf{Selected Bond Distances}}\\&\\
    &&Cr1-F&Cr2-F&Cr3-F& Cr4-F\\
    &&&\\
    $l$ $\times$ 2&&2.383(6) \AA&2.289(5) \AA&2.346(5) \AA&2.371(5) \AA  \\
    $m$ $\times$ 2&&2.028(5) \AA&2.045(5) \AA&2.019(5) \AA&2.022(5) \AA  \\
    $s$ $\times$ 2&&1.987(5) \AA&1.976(5) \AA&1.986(5) \AA&1.986(5) \AA  \\
    &&&\\
     \hline
   \end{tabular}
\end{table*}

\section{Experimental and computation section}
\label{sec:exo}

\subsection{Synthesis of $\mathrm{NaCrF_3}$}
\label{sec:syn}

Chromium              (II)              acetate              dihydrate
($\mathrm{Cr_2 (CH_3 CO_2)_4 (H_2 O)_2}$) (0.5g 1.33 mmol) and 2 mL of
degassed water is  loaded into a 85 mL polycarbonate  (PC) vial closed
with a septum under a constant  flow of Ar.  $\mathrm{NaHF_2}$ (0.45 g
5.45 mmol)  is dissolved in  10 mL deoxygenated  water in a  second PC
vial under  Ar by heating  to above 50$^\circ$C.  The  hot-solution of
$\mathrm{NaHF_2}$  is carefully  and  quickly injected  into the  vial
containing  $\mathrm{Cr_2 (CH_3  CO_2)_4  (H_2  O)_2}$ under  vigorous
stirring.   $\mathrm{NaCrF_3}$ precipitates  after  few seconds.   The
supernatant is decanted off and the  solid product is washed once with
2 mL 50:50 deoxygenated water  and methanol solution, and subsequently
with 5 mL deoxygenated methanol.  Finally, the product is vacuum-dried
overnight to yield air-stable $\mathrm{NaCrF_3}$.

\subsection{Computational simulations}
\label{sec:com}

For  the   structural  phase  model  of   $\mathrm{NaCrF_3}$,  density
functional theory (DFT)  was applied using the Vienna  {\em Ab initio}
Simulation  Package (\cite{VA1,VA2}),  with the  PBE general  gradient
approximation (GGA) \cite{PBE}.   The cutoff energy of  the plane wave
basis set  expansion was set  to at least 450  eV. The density  of $k$
points was determined  by a maximum of 0.25  \AA$^{-1}$. The structure
was relaxed  with remaining  forces below 0.05  eV/\AA\ using  a quasi
Newton method.

\subsection{Synchrotron X-ray diffraction}
\label{sec:sc}

High-  resolution  synchrotron  powder  X-ray  diffraction  (HR-SPXRD)
experiments  were   conducted  at   ID22  beamline  of   the  European
Synchrotron (ESRF),  Grenoble, France  where the  diffraction patterns
were recorded  using a  wavelength of $\lambda=0.40013$~\AA\;  at room
temperature.   The crystal  structure of  $\mathrm{NaCrF_3}$ has  been
refined using TOPAS  v5 (Bruker AXS) \cite{TOPAS}.   The initial model
was obtained by DFT minimisation of a symmetry free (space group $P1$)
triclinic model  based on the crystal  structure of $\mathrm{NaCuF_3}$
\cite{VOLL,KEISER}  with  Cr replacing  Cu.   This  model was  refined
against the HR-PXRD data to  obtain the correct lattice parameters and
crystallite size peak  broadening. The model was  then processed using
the  ADDSYMM   routine  in  PLATON  \cite{PLATON}   to  determine  the
crystallographic  symmetry.   The  new   model,  now  in  space  group
$P\bar{1}$,  was refined  against the  HR-SPXRD data.   Scale, lattice
parameters, 13 term Chebyshev polynomial background function, Gaussian
crystallite  size and  strain and  Lorentzian strain  broadening terms
(fundamental  parameters  peak   shape)  and  all  Na   and  F  atomic
coordinates and isotropic displacement  parameters were refined. Atoms
of the  same type (Na,  Cr and F)  were constrained to  have identical
isotropic thermal parameters.

\subsection{Magnetic characterization}
\label{sec:mc}

Magnetometry  experiments were  performed  on a  Quantum  Design 14  T
Physical Property Measurement System (PPMS).  Temperature dependent DC
magnetic susceptibility $\chi (T)$  measurements were conducted during
heating  from  4  to  300~K  in  zero-field-cooled  field-cooled  mode
(ZFC-FC).     The   magnetic    susceptibility   is    calculated   by
$\chi  =M/\mathrm{H}$ where  $M$  is the  magnetization  given in  emu
$\mathrm{mol^{-1}}$ and the magnetic  field $H=1$ T.  Isothermal field
dependent measurements ($M (H)$) were  collected at 2~K, and half-loop
isothermal measurements at 4, 12 and 23~K up to 14 T.

\subsection{Neutron Powder Diffraction}
\label{sec:npd}  Neutron   powder  diffraction  (NPD)   patterns  were
collected  at ISIS  Neutron and  Muon Source  (UK) by  using the  WISH
long-wavelength diffractometer \cite{WISH}.  The  sample was placed in
thin wall  vanadium container  (7~mm in diameter)  and cooled  down to
1.5~K.  The measurements were performed while heating from 1.5~K up to
127~K at  several temperature  steps. The raw  data was  integrated by
using the Mantid  suite \cite{MANTID} and analysed  using the Jana2006
software \cite{JANA}.   The structure  refinement was  performed using
data from the four detector banks with highest resolution.  The lowest
resolution bank was  discarded as the it contained  no information not
present  in  the  other  detector  banks.   The  background  (10  term
Chebyshev  polynomial),  peak-shape,  isotropic  thermal  displacement
parameters for each  element type, lattice parameters  and angles, and
scale  parameters   were  refined.    The  superspace   formalism  for
commensurate  magnetic moment  modulation  was used  for the  magnetic
structure description.  The magnetic form factor of $\mathrm{Cr^{2+}}$
was employed in  the refinements.  Spherical coordinates  were used to
refine the  magnetic moments.  The four  $\mathrm{Cr^{2+}}$ sites were
constrained to have  a single magnetic moment  magnitude. Polar angles
($\varphi_1$ and $\varphi_3$)  were refined for Cr1 and  Cr3, with the
polar angles of  Cr2 and Cr4 constrained to  values of 180+$\varphi_1$
and 180+$\varphi_3$  respectively.  Independent azimuthal  angles were
refined  for  all  Cr  sites.  These  constraints  are  summarised  in
\tref{t2}. At 17 and 19 K, the azimuthal angle of Cr1 and Cr3, and Cr2
and Cr4 were constrained to be equal. Also at 19~K the polar angle for
Cr1 and Cr3 was  fixed at values obtained at 17~K. This  is due to the
low  intensity   of  magnetic   Bragg  reflections  near   the  N\'eel
temperature and fit instability.

\begin{table*}[t!]
   \centering
   \caption{Parameters  used to  describe  the  magnetic structure  of
     $\mathrm{NaCrF_3}$  from   Rietveld  Refinements  at  1.5   K  in
     spherical coordinates with a modulation vector of $k=(1/2,1/2,0)$
     in  superspace group  $P\bar{1}(\alpha \beta  0)$.  The  magnetic
     moment was constrained to be  equal for all chromium atoms, while
     the polar  $\varphi$ and azimuthal $\vartheta$  angles were given
     degrees of freedom.  \\}
   \label{tab:t2}
   \begin{tabular}{c c c c c}
\hline\hline
       &&&&\\
     Atom Label      & Atom position & Magnetic moment & Polar angle & Azimuthal angle\\
                    &&&&\\
          Cr1	&(1/2 0 0) 		&	 M = 3.520(6)	& $\varphi_1 = 223.04(70)$	& $\vartheta_1 = 38.3(1.2)$\\
          Cr2	&(0 1/2 0)		&	 M = 3.520(6)	&$\varphi_1 + 180 = 43.04(70)$	&$\vartheta_2 = 128.9(1.0)$\\
          Cr3	&(1/2 0 1/2) 		&	 M = 3.520(6)	&$\varphi_3 = 223.04(70)$	&$\vartheta_3 = 55.5(1.0))$\\
          Cr4	&(0 1/2 1/2)		&	 M = 3.520(6)	&$\varphi_3 + 180 = 43.04(70)$	&$\vartheta_4 = 135.9(1.1)$\\
                    &&&&\\
     \hline
   \end{tabular}
 \end{table*}

\begin{figure*}[t!]
  \centering
 \includegraphics[scale=.7]{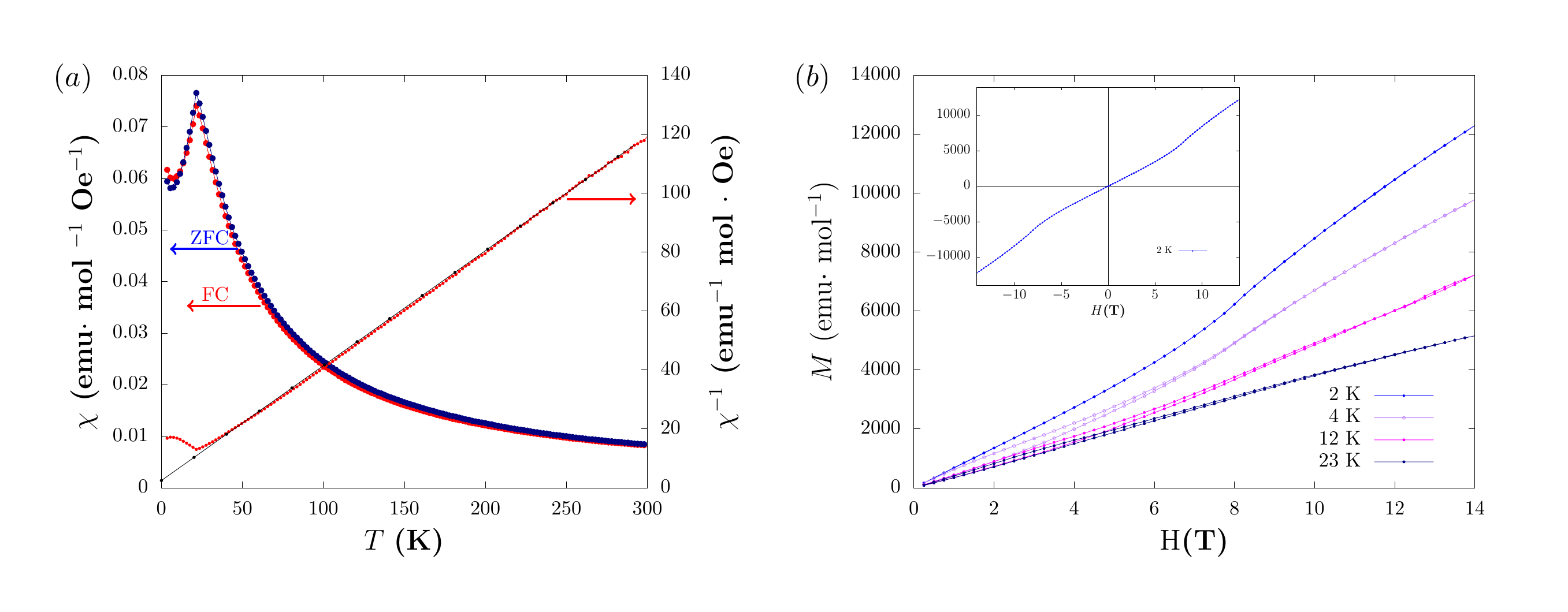}
 \caption{($a$)   ZFC-FC  temperature   dependency  of   the  magnetic
   susceptibility  measured $\chi  (T)$ at  $H=1$ T  (left axis),  and
   their  inverse $\chi^{-1}$  (right) with  the linear  regression at
   $\theta=-4$ K  .  ($b$)  Isothermal half-loop  magnetization curves
   magnetic field ($M (H)$) applied from 0  to 14 T and then back to 0
   T at 2, 4,  12 and 23 K.  The inset is the  full $M (H)$ hysteresis
   loop at 2 K to show the symmetry at the negative quadrant.}
  \label{fig:f2}
\end{figure*}

\section{Results}
\label{sec:results}

\subsection{Crystal structure determination}
\label{sec:sc}

To  the best  of our  knowledge,  no reliable  synthesis protocol  for
$\mathrm{NaCrF_3}$has  previously  been  described,  and  the  crystal
structure  of the  compound has  not  been described  in detail.   The
air-sensitivity  of $\mathrm{Cr^{2+}}$  is intrinsically  difficult to
combine with fluorine chemistry.  Conventional solid-state methods are
therefore  unsuitable,  so  we  developed a  novel  own  wet-chemistry
protocol.    Using   this  we   can   work   under  conditions   where
$\mathrm{Cr^{2+}}$   is  stable   and   obtain   pure,  single   phase
$\mathrm{NaCrF_3}$ in large quantities. We expect that other fluorides
can  be prepared  using the  same  approach. Results  of the  Rietveld
refinement against HR-PXRD  data are shown in  Figure \ref{fig:f1} and
Table \ref{tab:t1}.  The plot, fitting statistics and bond lengths and
angles obtained indicate that the model is an excellent representation
of the real structure.  \tref{t1} and \tref{s1} SI show the structural
parameters  and   atomic  coordinates,   as  obtained   from  Rietveld
refinements.

The    $\mathrm{Cr^{2+}}$   cations    occupy   four    non-equivalent
crystallographic sites.  Although the structure is triclinic, the cell
edges and  angles are close to  those of a tetragonal  unit cell.  The
corresponding   approximated    tetragonal   distortion,   $c/a=1.48$,
corresponds to  the (stretching) normal  mode $Q_2$ of  the octahedral
units  $\mathrm{CrF_6^{4-}}$.   \fref{f1}   ($b$)  shows  the  crystal
structure of  $\mathrm{NaCrF_3}$ with  vertex shared  octahedral units
(blue)  with  Na$^+$ ions  (red)  in  interstices.  We  calculate  the
octahedral      distortion     according      to     the      equation
$\Delta_d=  1/6 \sum_{n=1}^6|l_i-l_{av}|/l_{av}$  where $l_i$  are the
individual bond distances of the  octahedral unit, and $l_{av}$ is the
average bond distance.   \fref{f1b} ($a$) shows the $l$  and $s$ bonds
building  a tilted  $ls$-motif  connected through  the Cr-F-Cr  angles
$\xi_i^\circ$.  \fref{f1b} ($b$) shows  the $ls$- motif stacking along
[110], with the bonding-motif  rotated 90$^\circ$ (represented here as
blue and red planes to  indicate the 90$^\circ$ rotation), whereas the
$m$  bonds  propagate  above  and   below  the  $(001)$-plane  in  the
$[1\bar{1}0]$-direction.   The four  $\mathrm{CrF_6^{4-}}$ distortions
can be found  in \tref{t1}.  The non-equivalent  octahedra are sharply
tilted,   corresponding    to   the   Glazer    notation   $a^-b^-c^-$
\cite{GLAZER}.

\subsection{Magnetic characterization and Neutron diffraction studies}
\label{sec:MG}

DC  temperature dependent  magnetic  susceptibility  experiments on  a
polycrystalline sample of $\mathrm{NaCrF_3}$ between  4 and 300 K show
a  kink corresponding  to  the onset  of long-range  antiferromagnetic
ordering on reaching the N\'eel temperature at $T_N=21.3$ K, \fref{f2}
($a$).  Furthermore, an upswing at around 9~K reveals the emergence of
a  weak ferromagnetic  component at  lower temperature.   The Curie  -
Weiss (CW)  law is applicable  for the temperature  range 300 -  24 K.
The fit  to the inverse  susceptibility curves 1$/\chi$ show  a linear
behaviour    where    the    calculated   paramagnetic    moment    of
$\mu_{eff}=4.47 \mu_B$ is in good agreement with the theoretical value
of  the spin-only  configuration  $S=2$  for $\mathrm{Cr^{2+}}$.   The
Curie temperature is  $\theta=-4$ K measured under 1  T indicates just
weak    antiferromagnetic    interactions.   This    contrasts    with
$\mathrm{KCrF_3}$  which  displays   weak  ferromagnetic  interactions
$\theta=2.7 (1)$ K at 1  T \cite{JACS128} This suggests that reduction
of  the  ion size  at  the  $A$-site  is  of paramount  importance  in
finetuning the magnetic exchange interactions.

\begin{figure}[t!]
  \flushleft
 \includegraphics[scale=.65]{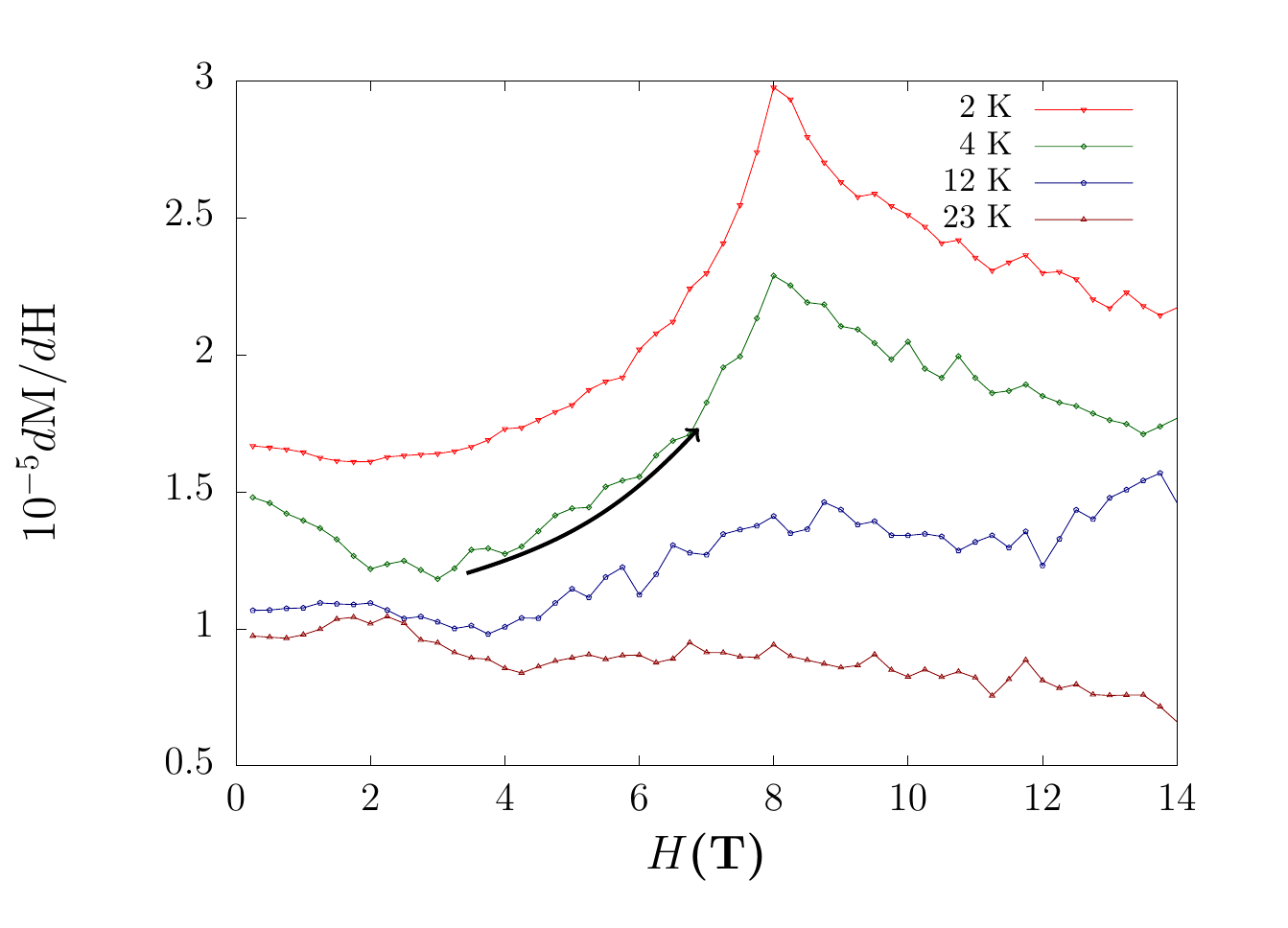}
 \caption{ First derivative $dM/dH$ of the isothermal half-loops at 2,
   4,  12,  and 23  K.   The  upswing  is represented  with  connected
   line-guides.   A  metamagnetic transition  occurs  at  8 T  in  the
   half-loops at 2  and 4 K.  Up and downswing  data are emphasized by
   arrows for the 4 T data.}
 \label{fig:f2d}
 %%%%%%%%%%%%%%%%%%%%%%%%%%%%%%%%%%%%%%%%%%%%%%%%%%%%%%%%%%%%%%%%%%%%%%
 %%  HF:  I dislike  that  the  green and  brown  1/X  curves are  not
 %% identical. FC vs ZFC should not matter. There should be no history
 %% of the field situation above TN. Also the paramagnetic moment will
 %% be different. Would uit be OK  just to keep the ZFC curve? Are the
 %% listed paramagnetic parameters referring  to green or brown curve?
 %% Which is which?
 %%%%%%%%%%%%%%%%%%%%%%%%%%%%%%%%%%%%%%%%%%%%%%%%%%%%%%%%%%%%%%%%%%%%%%
\end{figure}

Magnetic field  dependent isothermal  $M (H)$ half-loops  (forward and
reverse  field application)  for $\mathrm{NaCrF_3}$  are presented  in
\fref{f2} ($b$).  These  loops were measured at  2, 4, 12 and  23 K in
applied  magnetic fields  up to  14 T.   At 23  K the  half-loop shows
almost  linear   behaviour,  nevertheless  with  a   small  hysteresis
indicating the presence of  ferromagnetic interactions.  The half-loop
at  12  K  retains  the   hysteresis  with  additional  signatures  of
metamagnetic  transitions  identified  by  a  clear  S-shape  occuring
between 6 and 8 T.

The metamagnetic  transition becomes  more pronounced  with decreasing
temperature as observed at 4 and 2 K.  At 4 K the hysteresis is at its
widest.  However, as shown by complete isothermal loop in the inset to
\fref{f2} ($b$), there is no longer any hysteresis at 2 K.  This means
that  the  ferromagnetic components  are  suppressed  by lowering  the
temperature.   In   order  to  identify  the   point  of  metamagnetic
transition      we      calculated      the      first      derivative
$d\mathrm{M}/d\mathrm{H}$  of the  magnetization $M$  with respect  to
applied field $H$ as shown in  \ref{fig:f2d}.  An emergent peak at 8 T
is observed below $T_N$ with well-defined singularities at 4 and 2 K.

\begin{figure*}[t!]
  \centering
  \includegraphics[scale=.6]{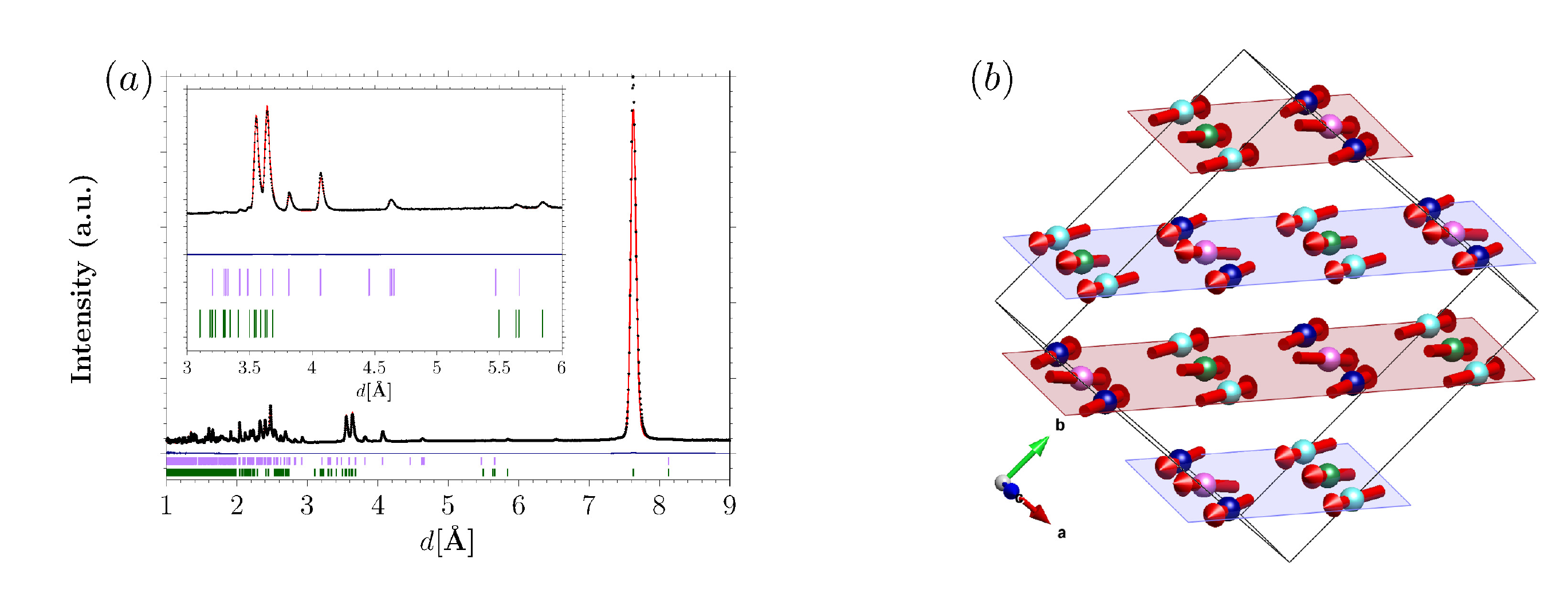}
  \caption{($a$)    Rietveld   refinements    of   NPD    dataset   of
    $\mathrm{NaCrF_3}$  at   1.5  K  from  detector   bank  2  (lowest
    resolution bank used in refinements)  showing the peak at 7.63 \AA
    $ $ with the inlet showing  small peaks of the magnetic phase. The
    purple  and green  tics  correspond to  the  crystal and  magnetic
    phase,    respectively.      ($b$)    Magnetic     structure    of
    $\mathrm{NaCrF_3}$  in the  [1,-1,0]-direction. The  anti-parallel
    alignment of  the spins is  represented by the  blue-red sequence.
    Blue, cyan, purple and green atoms correspond to Cr1, Cr2, Cr3 and
    Cr4 respectively.}
  \label{fig:f3}
\end{figure*}

The derived synthesis protocol made it possible to prepare large scale
samples with high  purity and crystallinity, well  suited for detailed
neutron diffraction studies.  We  conducted powder neutron diffraction
experiments between 1.5 and 127 K to study the structural and magnetic
changes in  $\mathrm{NaCrF_3}$ above  and below the  Néel temperature.
Visual  inspection  of  the  neutron diffraction  patterns  reveals  a
transition originating  from the ordering  of magnetic moments  in the
proximity of  the Néel temperature,  e.g.  a strong reflection  due to
magnetic  ordering occurs  at $d  = 7.63$  \AA, \fref{f3}  ($a$).  The
additional  magnetic  reflections were  indexed  in  a supercell  with
doubled $a$-  and $b$-unit cell  parameters ($2a\times 2b  \times c$),
corresponding  to   a  propagation   vector  of   $k=(1/2,1/2,0)$  for
modulation  of  the  magnetic  structure.  To  describe  the  magnetic
structure in detail, we use  magnetic superspace group formalism.  The
magnetic   structure   is   described    in   the   superspace   group
$P\bar{1}(\alpha  \beta  0)$  with a  commensurate  modulation  vector
$(1/2,1/2,0)$.   There  are no  symmetry  driven  restrictions on  the
magnetic moment components  for any of the 4 positions  occupied by Cr
atoms.

%%%%%%%%%%%%%%%%%%%%%%%%%%%%%%%%%%%%%%%%%%%%%%%%%%%%%%%%%%%%%%%%%%%%%%
$\mathrm{NaCrF_3}$   adopts   a  canted   $A$-type   antiferromagnetic
structure where  chromium has  an ordered magnetic  moment of  $\mu$ =
3.520(6) $\mu_B$ at  1.5 K, \fref{f3} ($b$).  The  magnetic moments of
chromium    atoms    are     ferromagnetically    ordered    in    the
$(1\bar{1}0)$-planes, i.e. along [110]  and [001].  We observe canting
in the $(1\bar{1}0)$-plane. This cancels  out within the magnetic unit
cell due to  AFM stacking along $[1\bar{1}0]$, which are  shown by red
and blue colored planes in \fref{f3}.  In the triclinic structure, the
magnetic moments of  chromium atoms point almost  directly through the
middle of the edge between the  equatorial and axial fluorine atoms of
the  JT  distorted   $\mathrm{CrF_6}$  octahedra.   Consequently,  the
magnetic moments  forming chains in the  $[11\bar{1}]$-direction.  The
canted  A-type antiferromagnetic  structure is  in agreement  with the
structural $ls$-motif  corresponding to ferromagnetic  interactions in
the    $(1\bar{1}0)$-plane    and    antiferromagnetic    interactions
perpendicular  to it.  The $a^-b^-c^-$  tilts  reduces the  $3d –  2p$
overlap  and  weaken thereby  the  superexchange  interactions in  the
presence of Na$^+$ ions.
%%%%%%%%%%%%%%%%%%%%%%%%%%%%%%%%%%%%%%%%%%%%%%%%%%%%%%%%%%%%%%%%%%%%%%

When the direction of the magnetic  moments of the four chromium sites
were  constrained  to be  either  parallel  or anti-parallel,  several
weaker  reflections   originating  from  magnetic  ordering   at  i.e.
$ d  = 5.64$ and  $5.86 $  \AA $ $  were not correctly  accounted for.
Therefore, we  applied a slightly  more complex set of  constraints to
the magnetic moment components.  \tref{t2} presents the minimal set of
magnetic  structure  parameters and  the  constraints  applied in  the
refinement. The  presence of the two  reflections (at $ d  = 5.64$ and
$5.86  $) clearly  shows that  the four  chromium sites  have slightly
different canting of their magnetic  moments.  These subtle aspects of
the magnetic structure could only be described due the high resolution
and excellent  signal to noise  ratio of the neutron  diffraction data
obtained  from  the  WISH  instrument  at  ISIS  (UK).   The  magnetic
structure of $\mathrm{NaCrF_3}$ is accurately  described at 1.5 K, and
details are given in \tref{ST5}.

\begin{figure}[t!]
  \flushleft \includegraphics[scale=.8]{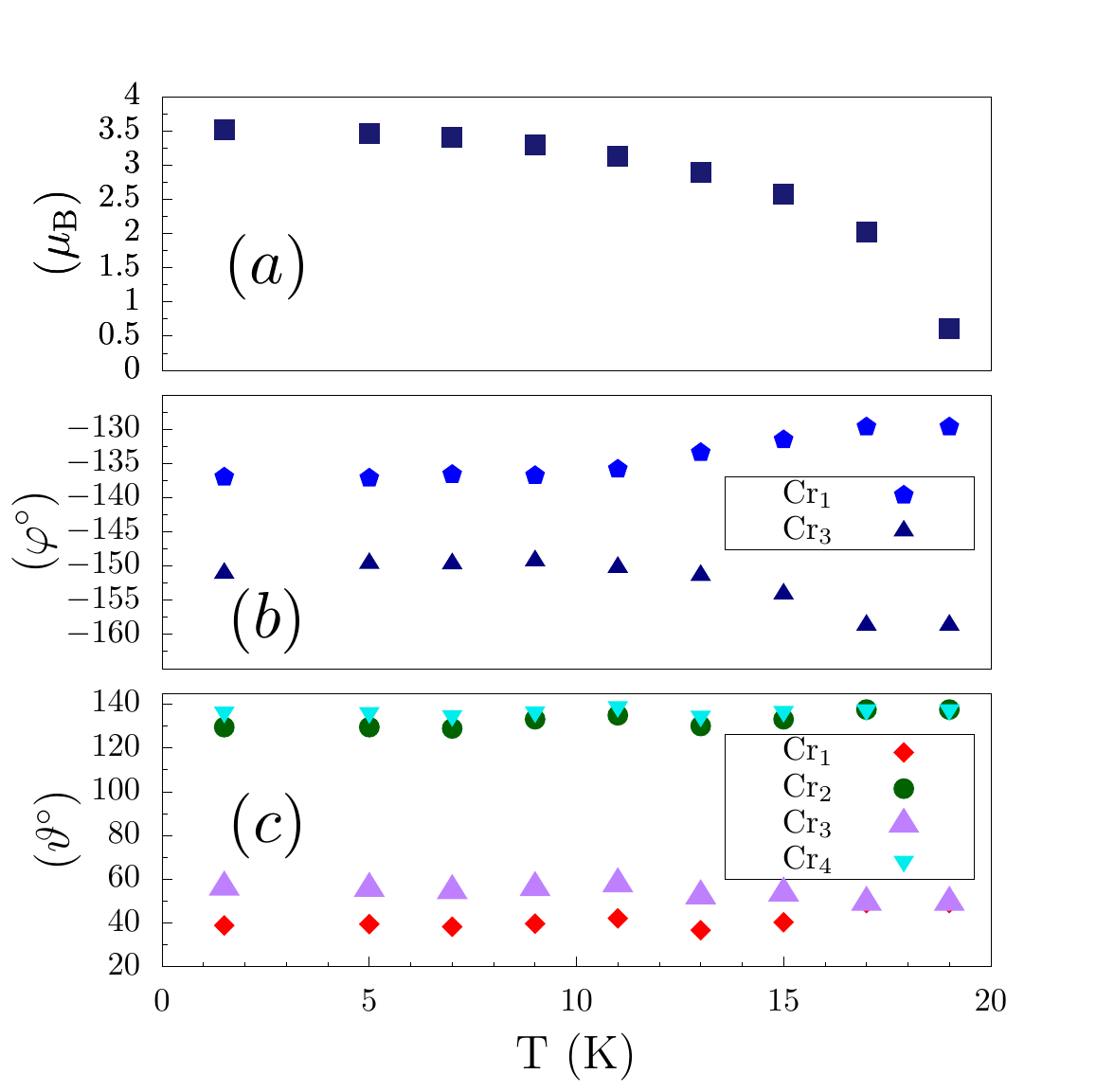}
  \caption{($a$)   Magnetic  moment   of  the   chromium  cations   in
    $\mathrm{NaCrF_3}$ determined by neutron diffraction as a function
    of  temperature.  Temperature  evolution  of the  ($b$) polar  and
    ($c$) azimuthal angles.  Constraints are described in \tref{t2}. }
	\label{fig:f6}
\end{figure}

The evolution of the magnetic  structure was further studied below the
N\'eel temperature ($T_N$ = 21.3~K).  In accordance with the spin only
approximation ($\mu_{eff}$ = 4.47~$\mu_B$  in the paramagnetic regime,
see above), the ordered magnetic moment of chromium is 3.520(6)$\mu_B$
at  1.5 K.   The slightly  lower  experimental value  compared to  the
theoretical value (of 4~$\mu_B$) is attributed to hybridization in the
chemical  bonding which  effectively reduces  the number  of electrons
contributing  to the  magnetic  moment.  The  ordered magnetic  moment
steadily  decreases  from  $\mu$  =  3.520(6)~$\mu_B$  at  1.5~K  with
increasing temperature  up to the Néel  temperature at 21 K  where the
magnetic ordering disappears (see Fig. \ref{fig:f6}($a$)).

The polar angle difference between the magnetic moments of Cr1 and Cr3
is fairly constant (see Fig.   \ref{fig:f6}($b,c$)).  The two pairs of
azimuthal angles  (Cr1, Cr3) and  (Cr2, Cr4) show similar  values, but
cannot  be   constrained  to   become  equal  without   worsening  the
fit. However,  at 17 and 19  K the azimuthal angle  $\vartheta$ of Cr1
and Cr3, and  Cr2 and Cr4 could successfully be  constrained.  For the
at 19~K data  all angle values were frozen at  values obtained at 17~K
due to  fit instability.  The  antiferromagnetic ordering at  the Néel
temperature is  associated with  a significant thermal  contraction of
the  lattice upon  cooling, \fref{f4}.   At the  ordering temperature,
changes in  the tilting of the  octahedra is revealed by  the analyzed
changes in the perovskite bond  angles.  These observations indicate a
clear magnetostructural coupling in $\mathrm{NaCrF_3}$.

\begin{figure}[t!]
 \includegraphics[scale=.75]{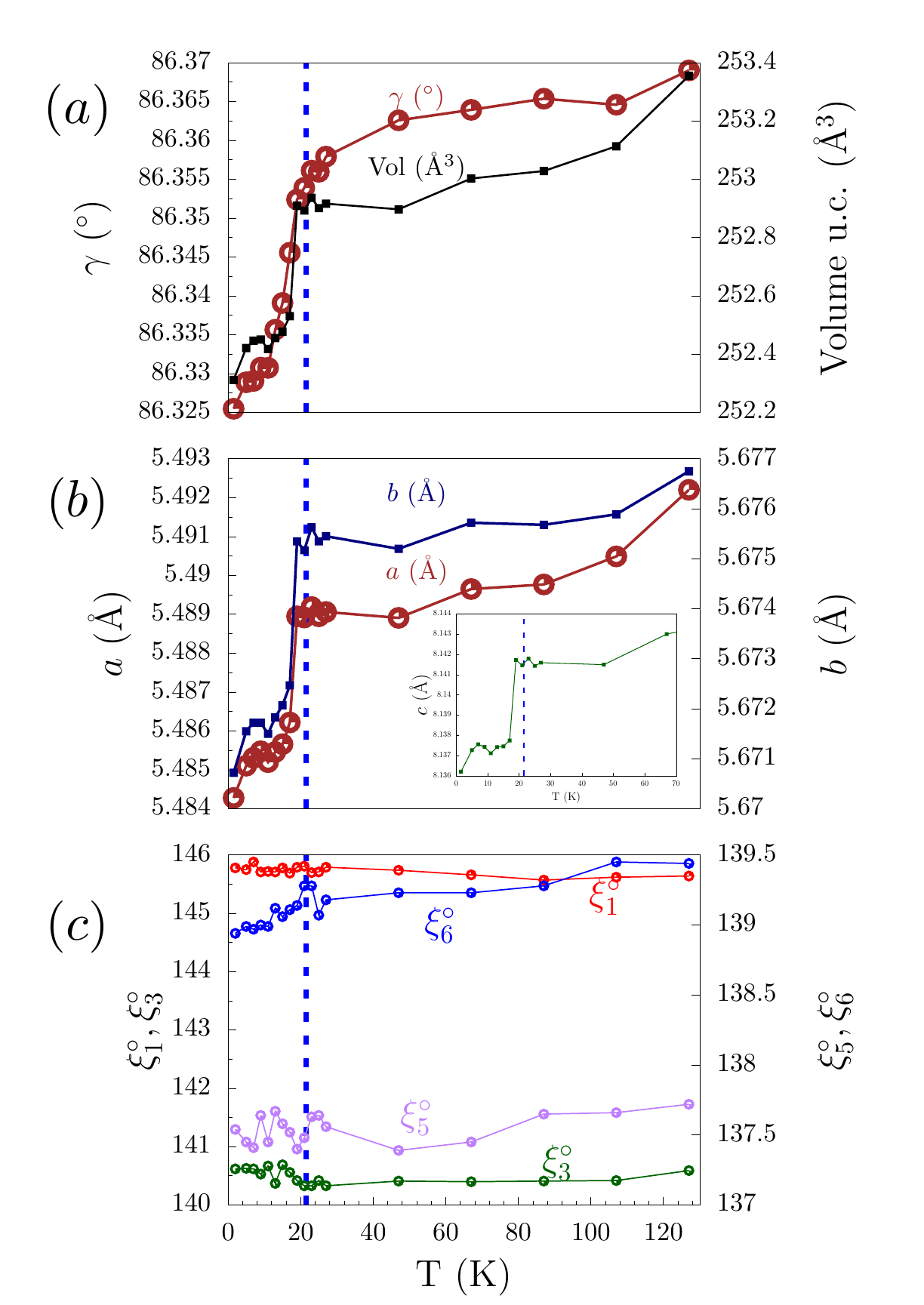}
 \caption{  ($a$)  Temperature  dependence of  unit  cells  dimensions
   ($a,  b,  \gamma,$ V)  and  $\xi^\circ$  angles. ($b$)  Temperature
   dependence  of the  unit cell  lattice parameters  ($c$) Perovskite
   angles  composing  the   canted  $ls$-motif  $\mathrm{NaCrF_3}$  as
   function    of   temperature:    $\xi^\circ_1=\mathrm{Cr1-F1-Cr2}$,
   $\xi^\circ_6=\mathrm{Cr2-F6-Cr4}$,
   $\xi^\circ_3=\mathrm{Cr4-F3-Cr3}$,
   $\xi^\circ_5=\mathrm{Cr3-F5-Cr1}$. Vertical dashed line at $T_N$ to
   emphasize the place where the magnetic long range order sets in.}
  \label{fig:f4}
\end{figure}

\section{Discussion}
\label{sec:disc}

The reliable new synthesis route  for $\mathrm{NaCrF_3}$ allowed us to
undertake a  detailed study of  its structure and  magnetic properties
for  the  first  time.    The  JT-active  ions  $\mathrm{Cr^{2+}}$  of
$\mathrm{NaCrF_3}$  occupy four  non-identical crystallographic  sites
with different  octahedral distortions.   Our results  demonstrate the
importance of the ion size at the $A$-site in tuning the properties of
the JT-active  $B$-site ions.   $A$-site dependent  physical phenomena
have previously  been observed  in the  $\mathrm{d^{4}}$ isoelectronic
low-dimensional manganese (III) fluoroperovskites, where variarions in
the $A$-site ion size give rise to rich and interesting phase diagrams
under external stimuli.

A  significant  feature  of  $\mathrm{NaCrF_3}$  is  its  metamagnetic
signatures below $T_N$ under  field dependent measurements in addition
to weak residual  ferromagnetic interactions at 23 K.  The presence of
metamagnetism in  $\mathrm{NaCrF_3}$ resembles  in some  aspects other
known  systems  with  exotic   properties  (See  Ref.   \cite{PRB100},
\cite{PRB96}, \cite{PRB98}).  We believe that this behavior is related
to correlations between the orbital structure and magnetic ordering as
discussed by Kugel and  Khomskii \cite{KK}.  The temperature dependent
NPD  data reveals  a  smooth  decrease in  the  unit  cell volume  and
$\gamma$  angle   above  $T_N$,  with   a  rapid  collapse   at  lower
temperatures \fref{f4}  ($a$).  One would expect  that the $\xi^\circ$
angle  would reduce  for all  four $\mathrm{Cr^{2+}}$  sites, however,
they   follow    independent   patterns   as   shown    in   \fref{f4}
($b$).  $\xi^\circ_6$ displays  a slight  decrease upon  cooling while
$\xi^\circ_1$ increases.  Perovskite angle reduction further decreases
the  orbital  overlap,  weakening   the  magnetic  interactions  while
reinforcing Cr-to-Cr  interactions.  The  refined magnetic  moments of
$\mathrm{Cr^{2+}}$ ions  in $\mathrm{NaCrF_3}$  are in  agreement with
NPD     studies    on     $\mathrm{KCrF_3}$    by     Xiao    \emph{et
  al.}\cite{PRB82}.   Compared  to   other  sodium   transition  metal
fluoroperovskites, $\mathrm{NaCrF_3}$  deviates from the  family trend
by  displaying a  canted $A-$type  magnetic ordering  compared to  the
$G$-types   found   in   $\mathrm{NaNiF_3}$   and   $\mathrm{NaCoF_3}$
\cite{NNiF,NCoF}.

To further investigate  the role of the  $A$-site in $A\mathrm{CrF_3}$
we report elsewhere the use of UV-vis spectroscopy along with magnetic
characterization studies to compare  the local electronic structure of
the JT-systems $\mathrm{KCrF_3}$ and  $\mathrm{NaCrF_3}$ as a function
of  temperature and  magnetic  field  \cite{MyPRB}.  Such  experiments
could  provide  more  detailed  information on  the  strength  of  the
JT-distortions and  be used to  assess OO-melting points  in JT-active
fluorides.

\section{Conclusions}
\label{sec:con}

This  work  provides  compelling  evidence of  the  existence  of  the
JT-active  compound $\mathrm{NaCrF_3}$,  and describes  its structural
and magnetic properties.  The successful development of a reliable and
reproducible synthesis  route, provided  the required  materials basis
for shedding  more light  on the  properties of  the $A\mathrm{CrF_3}$
family,  which previously  proved elusive  owing to  the air-sensitive
chemistry of  $\mathrm{Cr^{2+}}$.  The  structural and  magnetic phase
diagram  of  $\mathrm{NaCrF_3}$  is  much  simpler  than  the  diverse
situation observed for $\mathrm{KCrF_3}$  at low temperature.  This is
due  to  the  smaller  $A$-ion  size  causing  the  $\mathrm{NaCrF_3}$
structure  to  adopt  the  low  symmetry  space  group  $P\bar{1}$  at
relatively   high  temperature.    The  low   symmetry  structure   is
responsible    for    the    unusual    metamagnetic    behavior    of
$\mathrm{NaCrF_3}$, which can be clearly  linked to variations in both
the crystal  structure (perovskite angles and  lattice parameters) and
the magnetic  structure (polar  and azimuthal  angles of  the magnetic
moments),  observed in  the variable  temperature NPD  data.  The  new
synthesis  protocol opens  up  the possibility  of preparing  numerous
novel  stoichiometric compounds  by tuning  the $A$  and $B$  sites in
fluoroperovskites,  which  in  turn  may reveal  new  and  interesting
physical properties.

\section{Aknowledgements}
\label{sec:AK}

We  thank  Serena Margadonna  (Swansea  University,  Swansea, UK)  for
granted financing  support by  the Norwegian Research  Council (Norges
Forskningsr\aa d NFR) project 214260.  The U.K. Science and Technology
Facilities Council  (STFC) is thanked  for allocating beamtime  at the
ISIS  Facility.  We  also  thank  Pascal Manuel  for  help during  the
experiment.   We aknowldge  CRISMAT laboratory  (Caen France)  for the
magnetization measuraments  up to  14 T and  Fabien Veillon  and Bruno
Gonano  for technical  and analysis  help. We  thank Susmit  Kumar for
discussions.

\bibliographystyle{apssamp}
\bibliography{ACrF}

\newpage
 \begin{table*}[h!]
  \centering
  \caption{Atomic positions of $\mathrm{NaCrF_3}$ from HR-SPXRD Rietveld refinement. See \tref{t1} for crystal structure details. }
  \label{tab:s1}
  \begin{tabular}{l c c c c c c}
         &&&&&&\\
    \hline\hline
         &&&&&&\\
    Atom & Multiplicity & x & y & z & Occ & B$_{iso}$\\
         &&&&&&\\
    Na1&  2& 0.5062(8)& 0.5511(7)& 0.2370(5)& 1& 1.53(7)\\
    Na2&  2& 0.9765(8)& 0.0563(7)& 0.2603(5)& 1& 1.53(7)\\
    Cr1&  1& 0.5& 0& 0& 1& 0.79(3)\\
    Cr2&  1& 0& 0.5& 0& 1& 0.79(3)\\
    Cr3&  1& 0.5& 0& 0.5& 1& 0.79(3)\\
    Cr4&  1& 0& 0.5& 0.5& 1& 0.79(3)\\
    F1 &  2& 0.6788(9)& 0.2832(9)& 0.0558(6)& 1& 0.98(5)\\
    F2 &  2& 0.2019(9)& 0.2062(8)& 0.9260(6)& 1& 0.98(5)\\
    F3 &  2& 0.7153(9)& 0.3238(8)& 0.4292(6)& 1& 0.98(5)\\
    F4 &  2& 0.1887(10)& 0.1902(9)& 0.5503(7)& 1& 0.98(5)\\
    F5 &  2& 0.3801(9)& 0.9405(8)& 0.2724(7)& 1& 0.98(5)\\
    F6 &  2& 0.1204(9)& 0.4380(8)& 0.2272(7)& 1& 0.98(5)\\
         &&&&&&\\
    \hline
  \end{tabular}
\end{table*}

 \begin{table*}[h!]
   \centering
   \caption{Selected bond angels in $\mathrm{NaCrF_3}$ from HR-SPXRD Rietveld refinements}
   \label{tab:BANCrF}
   \begin{tabular}{c c c c c}
                    &&&&\\
\hline\hline
                    &&&&\\
     F$_i$-Cr-F$_j$ &Cr1 $_{(i=2,5,5;j=1,2,1)}$ |& Cr2 $_{(i=2,1,1;j=6,2,6)}$|& Cr3 $_{(i=4,3,3;j=5,4,5)}$|& Cr4 $_{(i=4,6,6;j=3,4,5)}$\\
                    &&&&\\
                    &90.8(2)&89.1(2)&88.5(2)&89.2(2)\\
                    &96.17(19)&91.72(19)&95.72(19)&94.00(19)\\
                    &94.61(19)&92.95(19)&94.19(19)&84.50(19)\\
     F$_i$-Cr-F$_i$&180.00(6) &179.999(7)&180.00(7)&180.00(9)\\
                    &&&&\\
     \hline
   \end{tabular}
 \end{table*}

\begin{table*}[th!]
	\centering
	\caption{Structural  parameters from  Rietveld  refinement of  NPD
		dataset of $\mathrm{NaCrF_3}$ at 1.5~K. $l,m$ and $s$
		are long, medium and short bond distances, respectively.}
	\label{tab:ST5}
	\begin{tabular}{l c c c c c}
		&&&&&\\
		\hline\hline
		&&&&&\\
		\textbf{Space Group}: & $P\bar{1}$&&&&\\
		$a$ &5.48428(11) \AA&&&&$\Delta d_{Cr1}=79.90\times 10^{-4}$ \AA \\
		$b$ &5.67072(12) \AA&&\textbf{Octahedral distortions}:&&$\Delta d_{Cr2}=54.68\times 10^{-4}$ \AA \\
		$c$ &8.13620(15) \AA&&&&$\Delta d_{Cr3}=71.39\times 10^{-4}$ \AA \\
		$\alpha$ & 90.3860(10)$^\circ$&&&&$\Delta d_{Cr4}=75.07\times 10^{-4}$ \AA\\
		$\beta$ &  92.2816(8)$^\circ$&&&\\
		$\gamma$& 86.3255(8)$^\circ$&&&\\
		$V$& 252.312(9) \AA$^3$&&&\\
		&&&&\\
		$R_{wp}$& 2.37 &&$R_{p} $&1.72 &\\
		N$^\circ$ of independent parameters &111&&&\\
		Restrains, constrains&0, 5&&&\\
		Rigid bodies        &0&&&\\
		$Z$&4&&&\\
		&\multicolumn{4}{c}{\textbf{Selected Bond Distances}}\\&\\
		&&Cr1-F&Cr2-F&Cr3-F& Cr4-F\\
		&&&\\
		$l$ $\times$ 2&&2.3807(19) \AA&2.2835(18) \AA&2.3480(17) \AA&2.3489(18) \AA  \\
		$m$ $\times$ 2&&2.0243(16) \AA&2.0550(16) \AA&2.0332(18) \AA&2.0131(17) \AA  \\
		$s$ $\times$ 2&&1.9727(17) \AA&1.9931(18) \AA&1.9808(18) \AA&1.9714(17) \AA  \\
		&&&\\
		\hline
	\end{tabular}
\end{table*}
%%%%%new table
 \begin{table*}[h!]
  \centering
  \caption{Atomic positions of $\mathrm{NaCrF_3}$ from NPD Rietveld refinements at 1.5 K. See \tref{ST5} for crystal structure details. }
  \label{tab:sm1}
  \begin{tabular}{l c c c c c c}
         &&&&&&\\
    \hline\hline
         &&&&&&\\
    Atom & Multiplicity & x & y & z & Occ & U$_{iso}$\\
         &&&&&&\\
     Na1   &       2 &     0.5084(5)        &       0.5551(4)        &       0.2358(3)        &       1       &       0.0181(4)\\
  Na2   &       2 &     0.9718(5)        &       0.0597(4)        &       0.2621(3)        &       1       &       0.0181(4)\\
  Cr1   &       1 &     0.5                      &       0                                &       0                                &       1       &       0.0085(4)\\
  Cr2   &       1 &     0                                &       0.5                      &       0                                &       1       &       0.0085(4)\\
  Cr3   &       1 &     0.5                      &       0                                &       0.5                      &       1       &       0.0085(4)\\
  Cr4   &       1 &     0                                &       0.5                      &       0.5                      &       1       &       0.0085(4)\\
  F1     &       2       &       0.6791(3)        &       0.2814(3)        &       0.0585(2)        &       1       &       0.0151(3)\\
  F2     &       2       &       0.2023(3)        &       0.2052(3)        &       0.9229(2)        &       1       &       0.0151(3)\\
  F3     &       2       &       0.7161(3)        &       0.3249(3)        &       0.4261(2)        &       1       &       0.0151(3)\\
  F4     &       2       &       0.1854(4)        &       0.1900(3)        &       0.5525(2)        &       1       &       0.0151(3)\\
  F5     &       2       &       0.3750(3)        &       0.9405(2)        &       0.2727(2)        &       1       &       0.0151(3)\\
  F6     &       2       &       0.1252(3)        &       0.4358(3)        &       0.2297(2)        &       1       &       0.0151(3)\\
         &&&&&&\\
    \hline
  \end{tabular}
\end{table*}
%%%%%%end

 \begin{table*}[h!]
   \centering
   \caption{Magnetic parameters of $\mathrm{Cr^{2+}}$ ions in $\mathrm{NaCrF_3}$ from Rietveld refinements of PND as function of temperature.  \\}
   \label{tab:BANCrF}
   \begin{tabular}{c c c c c c c c c c}
                    &&&&\\
\hline\hline
                    &&&&\\
Temperature&$\mathrm{M}$ & $\varphi({\rm Cr1})$ & $\vartheta({\rm Cr1})$  & $\varphi({\rm Cr2})$ & $\vartheta({\rm Cr2})$ & $\varphi({\rm Cr3})$ & $\vartheta({\rm Cr3})$ & $\varphi({\rm Cr4})$ & $\vartheta({\rm Cr4})$ \\
                    &&&&\\
1.5	&		3.51969		&	-136.9916	&	38.94935	&	43.00841	&	129.5479	&	-151.1129	&	56.12197	&	28.88715	&	136.5042	\\
5  	&		3.46608		&	-137.1359	&	39.5289 	&	42.86414	&	129.5129	&	-149.6677	&	55.52545	&	30.33231	&	136.3085	\\
7  	&		3.40534		&	-136.6089	&	38.33702	&	43.39112	&	128.9114	&	-149.7235	&	54.45972	&	30.27652	&	134.9575	\\
9  	&		3.29101		&	-136.7846	&	39.77436	&	43.21539	&	133.1239	&	-149.2775	&	55.90062	&	30.72252	&	136.6471	\\
11 	&		3.12823		&	-135.8486	&	42.12699	&	44.15142	&	134.918 	&	-150.2494	&	57.66518	&	29.75058	&	139.0407	\\
13 	&		2.9001 		&	-133.409 	&	36.74886	&	46.591  	&	130.0525	&	-151.4175	&	52.1276 	&	28.58247	&	134.7129	\\
15 	&     2.57635   	&  -131.5435	&  40.35261	&  48.4565 	&	133.1635	&	-154.1384	&	53.42544	&	25.86157	&	136.7845	\\
17 	&     2.02352   	&  -129.6754	&  49.17003	&  50.32455	&	137.5972	&	-158.6962	&	49.17003	&	21.30376	&	137.5972	\\
19 	&     0.6124    	&  -129.6754	&  49.17003	&  50.32462	&	137.5972	&	-158.6962	&	49.17001	&	21.30381	&	137.5972	\\
                    &&&&\\
     \hline
   \end{tabular}
 \end{table*}

\end{document}